\documentclass[runningheads]{llncs}
\usepackage{graphicx}
\usepackage{booktabs}
\usepackage{subfigure}
\usepackage{amsmath}
\usepackage{outlines}
\usepackage{color}
\usepackage{soul}
\begin{document}

\title{Exploring how Component Factors and their Uncertainty Affect Judgements of Risk in Cyber-Security\thanks{Supported by EPSRC’s EP/P011918/1 grant and by the UK National Cyber Security Centre (NCSC).}}

\titlerunning{Exploring how Factors and Uncertainty Affect Judgements of Risk}

\author{Zack Ellerby \and
Josie McCulloch\and
Melanie Wilson\and
Christian Wagner}

\authorrunning{Z. Ellerby \textit{et al.}}

\institute{
Computer Science, University of Nottingham, UK,\\
\email{\{zack.ellerby,josie.mcculloch,\\melanie.wilson,christian.wagner\}@nottingham.ac.uk}}
%

\maketitle              

\begin{abstract}
Subjective judgements from experts provide essential information when assessing and modelling threats in respect to cyber-physical systems. For example, the vulnerability of individual system components can be described using multiple factors, such as complexity, technological maturity, and the availability of tools to aid an attack. Such information is useful for determining attack risk, but much of it is challenging to acquire automatically and instead must be collected through expert assessments. However, most experts inherently carry some degree of uncertainty in their assessments. For example, it is impossible to be certain precisely how many tools are available to aid an attack. Traditional methods of capturing subjective judgements through choices such as \emph{high}, \emph{medium} or \emph{low} do not enable experts to quantify their uncertainty. However, it is important to measure the range of uncertainty surrounding responses in order to appropriately inform system vulnerability analysis. We use a recently introduced interval-valued response-format to capture uncertainty in experts' judgements and employ inferential statistical approaches to analyse the data. We identify key attributes that contribute to hop vulnerability in cyber-systems and demonstrate the value of capturing the uncertainty around these attributes. We find that this uncertainty is not only predictive of uncertainty in the overall vulnerability of a given system component, but also significantly informs ratings of overall component vulnerability itself. We propose that these methods and associated insights can be employed in real world situations, including vulnerability assessments of  cyber-physical systems, which are becoming increasingly complex and integrated into society, making them particularly susceptible to uncertainty in assessment.

\end{abstract}

\section{Introduction\label{sec:introduction}}
Cyber-security professionals play a vital role in assessing and predicting vulnerabilities within cyber-physical systems, which often form part of an organisation's or state's critical digital infrastructure. As outsider threats become more prevalent and sophisticated, there is increasing pressure on experts to provide timely and comprehensive assessments within the context of the rapidly changing cyber-physical ecosystem. As cyber-systems increase in both ubiquity and complexity, methods to quantify and handle error in subjective measurements from experts need to be developed \cite{koubatis2005risk}. It has been demonstrated across many industry sectors that as complexity increases accurate risk assessment decreases \cite{gardner2009risk}.  Enabling the effective reconstruction of overall assessments from component and attribute ratings would streamline the process of updating overall system vulnerability assessments, in line with shifts in this ecosystem.

Both objective and subjective measures of risk provide useful information to aid decision making in vulnerability assessment \cite{choi2004risk,duan2018novel}. Several different methods can be used to assess vulnerability and risk in a cyber-security system, such as vulnerability scanning tools \cite{munir2013quantitative} or the Common Vulnerability Scoring System (CVSS) \cite{mell2007complete}, which gives qualitative severity ratings of \emph{low}, \emph{medium}, and \emph{high}, and CVSS Version 3, which extends the ratings to include \emph{none} and \emph{critical} \cite{cvssspec}. However, when using CVSS, the necessary information to complete the calculation may be missing. Hubbard and Seiersen \cite{hubbard2016measure} argue that assessing risk in terms of \emph{low}, \emph{medium}, and \emph{high} ratings is highly subjective and open to error. It is therefore suggested that cyber-security risk should be described quantitatively. This would help quantify what areas of risk are perceived as important to cyber-security professionals and, from that, move towards how those risks might correspond to the actually enacted attacks and their success or failure.

Assessment of risk or the likelihood of an attack are inherently uncertain  \cite{aven2009risk,black2008cyber}. Objective measures may carry uncertainty because the measures themselves are imprecisely defined  \cite{black2008cyber}. Subjective assessments (collected from experts) carry uncertainty because, for example, the experts are not familiar with the particular technology, there is inherent uncertainty caused by insufficient detail in the scenario, or due to individual personality \cite{kahneman1982judgment,miller2013towards,slovic2016perception}.

Between-expert uncertainty is often modelled implicitly, for example, through probability distributions \cite{fielder2017uncertainty} or uncertainty measures \cite{duan2018novel}. These methods model the between-expert uncertainty, but they do not capture within-expert uncertainty. Choi et al. \cite{choi2004risk} capture within-expert uncertainty by enabling experts to express knowledge and uncertainty through terms such as \emph{very small}, \emph{small} and \emph{large} and using fuzzy sets to represent the uncertainty of these words. However, this assumes that the experts share the same degree of uncertainty regarding the meanings of these terms. After capturing uncertainty, it may be modelled and handled through methods such as Dempster-Shafer evidence theory \cite{feng2011information,gao2008analysis} or fuzzy logic \cite{choi2004risk,linda2011fuzzy,sikos2018handling}.

\begin{figure}[t]
\centering
  \subfigure[]{\includegraphics[width=8cm]{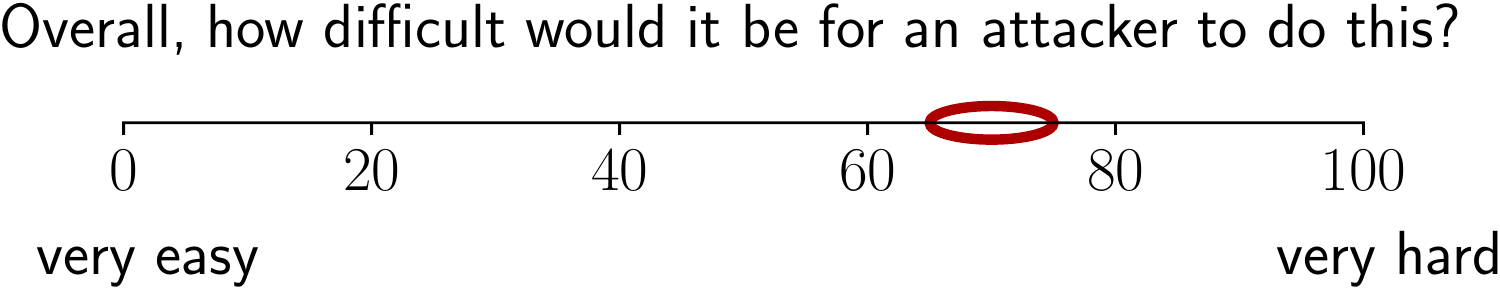}}\\\vspace{0.5cm}
  \subfigure[]{\includegraphics[width=8cm]{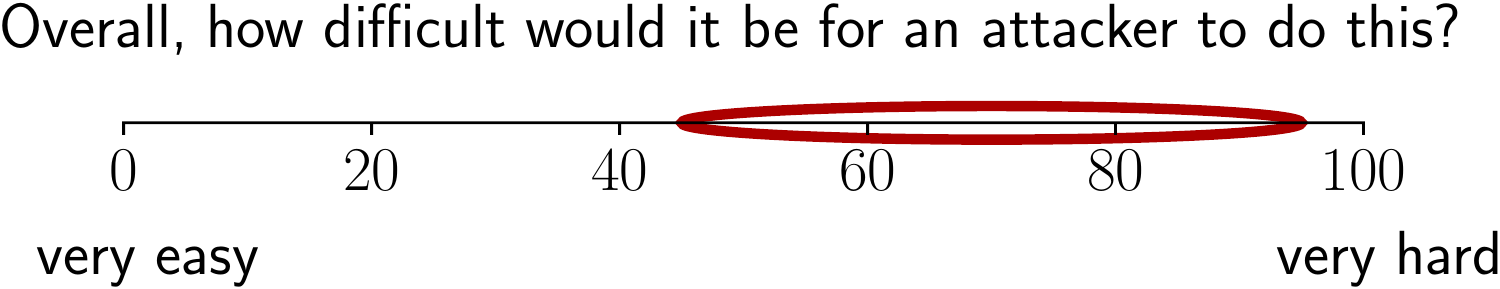}}
  \caption{Illustration of narrow (a) and wide (b) interval-valued responses capturing different degrees of uncertainty.\label{fig:responses}}
\end{figure}

We propose explicitly capturing uncertainty in experts' individual judgements using an interval-valued response format, as previously introduced in \cite{miller2016modelling}. Experts provide ratings along a continuous scale to quantify, for example, the perceived difficulty of an attack on a component. Using an interval captures both the experts' rating (position on the axis) and the degree of uncertainty associated with the response (width of the interval). Fig. \ref{fig:responses} shows an example of a narrow rating (slightly uncertain) and a wide rating (highly uncertain). In this manner, uncertainty is captured as an integral aspect of the judgement itself, through a coherent and intuitive response-format. The novel output of this paper is to show that an interval-valued response scale can be used to effectively capture uncertainty in expert judgements, and that this can be used to better predict both the magnitude and the uncertainty of risk in vulnerability assessments.

In this paper, we assess the importance of a variety of attributes in determining the overall vulnerability of components being attacked or evaded within a cyber-system; these include maturity of the technology and the frequency that a given attack is reported (the full set of attributes are listed in Tables \ref{tab:attributes_attack} and \ref{tab:attributes_evade}). We also capture the overall difficulty of attacking or evading each component. Our core aim is to understand how the component attributes contribute to the overall difficulty of an attack and, equally importantly, how uncertainty in component attributes affects not only the uncertainty in the overall vulnerability of a component, but also the overall difficulty itself. For example, experts might perceive an attack as more difficult if there are \emph{fewer} tools available to aid in this effort. However, experts might also perceive an attack as more difficult if they are \emph{uncertain} about the availability of tools. From this, we can learn what additional insight is gained by capturing uncertainty through interval-valued responses. Specifically, we wish to answer 
\begin{outline}
    \1 how is overall vulnerability of a component affected
      \2 by attribute ratings?
      \2 by uncertainty around attribute ratings?
    \1 how is uncertainty around overall vulnerability of a component affected
      \2 by attribute ratings?
      \2 by uncertainty around attribute ratings?
\end{outline}
We find that although cyber-security experts only assessed attributes that were previously deemed likely to be important to component vulnerability, only some of these attributes have a significant effect. Similarly, we discover that although uncertainty around some attributes affects the uncertainty around the component as a whole, this is not consistent for all attributes. We also find that the uncertainty around some attribute ratings makes a significant contribution to overall vulnerability ratings themselves.

\section{Methods\label{sec:methods}}
\subsection{Data Collection}
The data was collected from experts in CESG (Communications-Electronics Security Group), which was the information security arm of GCHQ (Government Communication Head Quarters) in the United Kingdom\footnote{CESG has since been replaced by the NCSC (National Cyber Security Centre).}. A total of 38 cyber-security experts at CESG assessed a range of components that are commonly encountered during a cyber-attack. They rated these on both overall difficulty to either attack or evade, as appropriate, and on several attributes that might affect this difficulty. Two types of components (also referred to as hops) were assessed, those that require an attacker to attack the component (referred to as \emph{attack}) and those that require the attacker to bypass the component (referred to as \emph{evade}). Examples of the hops assessed include \textit{bypass gateway content checker} and \textit{overcome client lockdown}, but any hop may be assessed using this method.

Tables \ref{tab:attributes_attack} and \ref{tab:attributes_evade} list the attack and evade attributes, respectively (variable notations are also provided, which are used in the next section). Attack hops are defined by seven attributes and evade hops by three attributes. In addition, both are described by their \emph{overall difficulty}. Our aim is to understand how the hop attributes and the uncertainty around these attributes relate to the perceived overall difficulty of attacking/evading the hop, and to the uncertainty around this difficulty. Experts were asked to provide interval-valued ratings to enable them to coherently express the uncertainty associated with their responses; these ratings were provided on a scale from 0 to 100.

The cyber-security components chosen for this study are designed to be representative of a mainstream government system, which would include assets at Business Impact Level 3 (BIL3) -- an intermediate category of impact \cite{cesg_2009}. CESG describes such a system as typically including remote site and mobile working, backed by core services and back-end office integrated systems, such as telemetry devices and associated systems used by the emergency services. Compromising assets at BIL3 might have large-scale negative effects, including but not limited to:  disruption to regional power supply, key local transport systems, emergency or other important local services for up to 24 hours, local loss of telecoms, risk to an individual's safety, damage to intelligence operations, hindrance to low level crime detection and prosecution, or financial loss to the UK government or a leading financial company in the order of millions (GBP) \cite{cesg_2009}. 

\begin{table}[t]
 \centering
 \caption{Attributes used to describe attack hops, the question used in the study, and the variable name used in the analysis.\label{tab:attributes_attack}}
 \def\arraystretch{1.5}
 \begin{tabular}{lp{2.9cm}p{7.6cm}}
  \toprule
  var & attribute & description \\
  \midrule
  $c$ & complexity &  How complex is the target component (e.g. in terms of size of code, number of sub-components)? \\
  $t$ & interaction & How much does the target component process/interact with any data input? \\
  $f$ & frequency & How often would you say this type of attack is reported in the public domain? \\
  $a$ & availability of tool & How likely is it that there will be a publicly available tool that could help with this attack? \\
  $d$ & inherent difficulty & How inherently difficult is this type of attack? (i.e. how technically demanding would it be to do from scratch, with no tools to help.) \\
  $r$ & maturity & How mature is this type of technology? \\
  $g$ & going unnoticed & How easy is it to carry this attack out without being noticed? \\
  \midrule
  $o$ & overall difficulty & Overall, how difficult would it be for an attacker to do this? \\
  \bottomrule
 \end{tabular}
\end{table}

\begin{table}[]
 \centering
 \caption{Attributes used to describe evade hops, the question used in the study, and the variable name used in the analysis.\label{tab:attributes_evade}}
 \def\arraystretch{1.5}
 \begin{tabular}{lp{2.9cm}p{7.6cm}}
  \toprule
  var & attribute & description \\
  \midrule
  $c$ & complexity & How complex is the job of providing this kind of defence? \\
  $a$ & availability of information & How likely is that there will be publicly available information that could help with evading defence? \\
  $r$ & maturity & How mature is this type of technology? \\
  \midrule
  $o$ & overall difficulty & Overall, how difficult would it be for an attacker to do this? \\
  \bottomrule
 \end{tabular}
\end{table}

\subsection{Analysis}
We use linear mixed effects modelling (an extension of linear regression) to determine the contribution of each of the hop attributes, as rated by experts, to overall hop difficulty. We also assess the contribution of the associated uncertainty in these ratings, captured through the interval-valued response-format. The midpoint ($m$) of the interval-valued response is used as a single-valued numeric rating of the attribute, and the width ($w$) of the response is used to represent the uncertainty around this rating.  Note, of course, that higher widths are only possible towards the centre of the scale. That is, as the midpoint approaches the edge of the scale, a wide interval cannot exist. Also, note that while experts provided ratings in the range $[0, 100]$, these data were standardised, through z-transformation, before entry into the model.

This approach estimates the contribution of each attribute's midpoint and width together upon the same outcome variable, in the form of $\beta$ weights. These variables are entered as fixed effects. The inclusion of random intercepts also allows the model to account for potential between expert and between hop differences in baseline ratings. In addition, this technique allows us to examine the combined effects of attribute rating and uncertainty (e.g. high certainty may have an opposite effect on overall difficulty when relating to a high or low attribute rating). We model this through the inclusion of two-way interaction terms ($m \cdot w$) pertaining to the midpoint and width of each attribute.

Four separate analyses are reported. These were conducted for the dependent variables of
\begin{itemize}
 \item attack hop overall difficulty rating (interval midpoint)
 \item attack hop overall uncertainty (interval width)
 \item evade hop overall difficulty rating (interval midpoint)
 \item evade hop overall uncertainty (interval width)
\end{itemize}
For each of these analyses, an initial model was created. These included fixed effects of all hop attribute ratings, all hop attribute widths and all two-way interactions, along with random intercepts for both expert and hop. Following this, a stepwise variable reduction process was applied to each model, in order to remove variables that were found not to significantly contribute to the respective outcome variable. $\beta$ weights of the variables retained into the final models were then interpreted as estimates of the (significant) contribution of each of these factors to the respective outcome variable.

Table \ref{tab:attributes_attack} lists the variables used to denote the attributes of attack hops. The sum of all simple effects for the attack hop attribute midpoints (ratings) is
\begin{equation}
 A^z_m = \beta^z_{1}x^{cm}_{i,j} + \beta^z_{2} x^{tm}_{i,j} + \beta^z_{3} x^{fm}_{i,j} + \beta^z_{4} x^{am}_{i,j}+ \beta^z_{5} x^{dm}_{i,j} + \beta^z_{6} x^{rm}_{i,j} + \beta^z_{7} x^{gm}_{i,j}
\end{equation}
where $\beta$ is the coefficient, $x^{cm}_{i,j}$ is the value $m$ (midpoint) of attribute $c$ (complexity) for $i$ (a given expert) and $j$ (a given hop), and $z$ reflects the model's outcome variable, which may be either $m$ (midpoints) or $w$ (widths) of the overall difficulty.

The sum of all simple effects for the attack hop attribute widths (uncertainty) is
\begin{equation}
 A^z_w = \beta^z_{8} x^{cw}_{i,j} + \beta^z_{9} x^{tw}_{i,j} + \beta^z_{10} x^{fw}_{i,j} + \beta^z_{11} x^{aw}_{i,j}+ \beta^z_{12} x^{dw}_{i,j} + \beta^z_{13} x^{rw}_{i,j} + \beta^z_{14} x^{gw}_{i,j}
\end{equation}
where $x^{cw}_{i,j}$ is the width $w$ of attribute $c$ (complexity) for $i$ (a given expert) and $j$ (a given hop).

The sum of the interactions between the midpoints and widths of the attack hop attributes is
\begin{align}
 A^z_{mw} = & \beta^z_{15} (x^{cm}_{i,j} \cdot x^{cw}_{i,j}) + 
 \beta^z_{16} (x^{tm}_{i,j} \cdot x^{tw}_{i,j}) + 
 \beta^z_{17} (x^{fm}_{i,j} \cdot x^{fw}_{i,j}) + 
 \beta^z_{18} (x^{am}_{i,j} \cdot x^{aw}_{i,j}) + \nonumber\\
 & \beta^z_{19} (x^{dm}_{i,j} \cdot x^{dw}_{i,j}) + 
 \beta^z_{20} (x^{rm}_{i,j}  \cdot x^{rw}_{i,j}) + 
 \beta^z_{21} (x^{gm}_{i,j} \cdot x^{gw}_{i,j})
\end{align}
Our initial model formula to explain the overall difficulty rating midpoints ($\gamma_{i,j}^{Aom}$) and widths ($\gamma_{i,j}^{Aow}$) of attack hops is then
\begin{equation}
 \gamma_{i,j}^{Aoz} = \beta^z_0 + A^z_m + A^z_w + A^z_{mw} + \mu^z_i + \mu^z_j + \epsilon^z_{i,j}
\end{equation}
where $z$ reflects the model's outcome variable, which may be either $m$ (midpoints) or $w$ (widths), for expert $i$ on hop $j$; $\beta_0$ denotes the fixed intercept; $\mu_i$ and $\mu_j$ denote respective random intercepts for expert and hop; and $\epsilon$ represents the error. The remaining $\beta$ terms (within $A_m$, $A_w$ and $A_{mw}$) denote the coefficients of the fixed effects of the hop attributes.

We perform likewise calculations for the evade hops (variables listed in Table \ref{tab:attributes_evade}).
The sum of effects for the midpoints of the evade hops is
\begin{equation}
 E^z_m = \beta^z_{1} x^{cm}_{i,j} + \beta^z_{2} x^{am}_{i,j} + \beta^z_{3} x^{rm}_{i,j},
\end{equation}
for the widths is 
\begin{equation}
 E^z_w = \beta^z_{4} x^{cw}_{i,j} + \beta^z_{5} x^{aw}_{i,j} + \beta^z_{6} x^{rw}_{i,j},
\end{equation}
and for the interactions is
\begin{equation}
 E^z_{mw} = \beta^z_{7} (x^{cm}_{i,j} \cdot x^{cw}_{i,j}) + \beta^z_{8} (x^{am}_{i,j}\cdot x^{aw}_{i,j}) + \beta^z_{9} (x^{rm}_{i,j}\cdot x^{rw}_{i,j})
\end{equation}
Our initial model formula to explain the overall difficulty ratings for midpoints ($\gamma_{i,j}^{Eom}$) and widths ($\gamma_{i,j}^{Eow}$) of evade hops is then
\begin{equation}
 \gamma_{i,j}^{Eoz} = \beta^z_0 + E^z_m + E^z_w + E^z_{mw} + \mu^z_i + \mu^z_j + \epsilon^z_{i,j}.
\end{equation}

Each of the initial models, as presented above, was then subjected to a backwards stepwise variable elimination procedure. During this, fixed effects were iteratively assessed and those that did not significantly contribute to the overall model were removed. Specifically, this process began by selection, from the pool of all non-significant fixed effects, of the effect with the \emph{t}-statistic closest to zero. This variable was then removed, and the resulting model directly compared with the preceding one, using the Theoretical Likelihood Ratio test. This was implemented through the MATLAB $fitlme$ and $compare$ functions. If the benefit of retaining the variable in question was calculated to be non-significant, then the model with the lower Bayesian Information Criterion (BIC) was retained into the next iteration. This procedure continued until a final model was determined, within which all fixed effects were statistically significant.

\section{Results\label{sec:results}}
\subsection{Attack Hops}

Table \ref{tab:attack_m} shows all effects retained in the final model with the outcome variable of overall attack hop difficulty (midpoints), following the stepwise variable reduction process. These results indicate that a number of factors make a substantial contribution. Attacks were rated less difficult if they are frequently reported or have a large availability of tools. By contrast, attacks were rated as more difficult if they have a greater inherent difficulty or relate to more mature technologies. Attacks were also rated as more difficult when technological maturity was uncertain and, perhaps surprisingly, when easier to go unnoticed. The latter might relate to some underlying factor - for instance, some attacks may be difficult to conduct, but also difficult to detect.  The significant interaction term ($m \cdot w$) indicates a combined effect of reported tool availability and uncertainty around this. This likely reflects that a hop is rated as being more difficult to attack when experts are certain about availability being low, but less difficult when experts are certain about availability being high. Unsurprisingly, the inherent difficulty rating was found to have the most robust effect.

\begin{table}
    \centering
    \caption{Results showing significant effects of hop attribute midpoints ($m$), widths ($w$) and two-way interactions ($m \cdot w$) on midpoints of overall attack hop difficulty ratings.\label{tab:attack_m}}
    \begin{tabular}{lrrrr}
    \toprule
    Fixed Effects Estimates & $\beta$ & SE & $t$ & $p$ \\
     \midrule
        Intercept : $(_0)$                                                         & .012  & .066 & .175   & .861   \\
        Frequency $m$ : $(x^{fm}_{i,j})$                                        & -.223 & .044 & -5.065 & $<.001$ \\
        Availability Tool $m$ : $(x^{am}_{i,j})$                                     & -.201 & .044 & -4.574 & $<.001$ \\
        Inherent Difficulty $m$ : $(x^{dm}_{i,j})$                              & .357  & .030 & 11.890 & $<.001$ \\
        Maturity $m$ : $(x^{rm}_{i,j})$                                         & .126  & .030 & 4.159  & $<.001$ \\
        Going Unnoticed. $m$ : $(x^{gm}_{i,j})$                                 & .142  & .027 & 5.194  & $<.001$ \\
        Maturity $w$ : $(x^{rw}_{i,j})$                                         & .071  & .027 & 2.612  & .009    \\
        Availability Tool $m \cdot w$ : $(x^{am}_{i,j} \cdot x^{aw}_{i,j})$ & .077  & .036 & 2.168  & .031    \\
    \midrule
         Random Effects Estimates & $\mu$ \\
    \midrule
         Expert intercept $(_i)$ & .183\\
         Hop intercept $(_j)$    & .204\\
    \midrule
      Residual $\epsilon_{i,j}$ & .502 \\
    \bottomrule
    \multicolumn{5}{r}{N = 532, DF = 524, AIC = 896.7, BIC=943.6}
    \end{tabular}
\end{table}

Table \ref{tab:attack_w} shows all effects retained in the final model with the outcome variable of uncertainty surrounding overall attack hop difficulty (widths). Even more factors were retained in this final model. Results indicated that experts were more certain about the vulnerability of hops on which attacks were reported more frequently -– likely due to familiarity. They were also more certain regarding hops that relate to mature technologies or when tool availability is low. By contrast, overall uncertainty significantly increased in line with attribute uncertainty for reported attack frequency, tool availability, inherent difficulty, and ease of going unnoticed. Two interaction terms were also retained, indicating that the effects of uncertainty around these attributes on overall uncertainty were significantly modulated by attribute rating, or vice versa. These can be interpreted together with main effects, depending upon direction. For example, in the case of ease of going unnoticed, there is a relatively large positive main effect of attribute uncertainty and a smaller, but significant, negative interaction term. This indicates that overall ratings were most uncertain when going unnoticed was considered difficult but uncertain. However, overall ratings were most certain when going unnoticed was considered difficult with certainty. The effect of attribute uncertainty was reduced, though still substantial, around hops rated as easier to go unnoticed when attacking. For maturity, however, there is a negative main effect of attribute rating and a negative interaction term, of comparable size. This indicates that overall uncertainty was greatest when maturity was rated low but uncertain. Also, while an increase in maturity rating tended to increase the certainty of overall ratings, this effect was driven by cases in which maturity was itself uncertain. Of all effects in this analysis, uncertainty surrounding the inherent difficulty rating was found to be the most robust.

\begin{table}[t]
    \centering
    \caption{Results showing significant effects of hop attribute midpoints (m), widths (w) and two-way interactions ($m \cdot w$) on widths of overall attack hop difficulty ratings.\label{tab:attack_w}}
    \begin{tabular}{lrrrr}
    \toprule
    Fixed Effects Estimates & $\beta$ & SE & $t$ & $p$ \\
     \midrule
        Intercept : $(_0)$                                                       & -.031 & .036 & -.857  & .392    \\
        Frequency $m$ : $(x^{fm}_{i,j})$                                       & -.116 & .045 & -2.614 & .009    \\
        Availability Tool $m$ : ($x^{am}_{i,j})$                               & .131  & .045 & 2.934  & .003    \\
        Maturity $m$ : $(x^{rm}_{i,j})$                                        & -.093 & .031 & -3.013 & .003    \\
        Frequency $w$ : $(x^{fw}_{i,j})$                                       & .141  & .035 & 4.034  & $<.001$ \\
        Availability Tool $w$ : $(x^{aw}_{i,j})$                               & .095  & .039 & 2.420  & .016    \\
        Inherent Difficulty $w$ : $(x^{dw}_{i,j})$                             & .406  & .037 & 10.959 & $<.001$ \\
        Going Unnoticed $w$ : $(x^{gw}_{i,j})$                                 & .268  & .036 & 7.399  & $<.001$ \\
        Maturity $m \cdot w$ : $(x^{rm}_{i,j} \cdot x^{rw}_{i,j})$         & -.122 & .035 & -3.484 & $<.001$ \\
        Going Unnoticed $m \cdot w$ : $(x^{gm}_{i,j} \cdot x^{gw}_{i,j})$  & -.080 & .035 & -2.270 & .024    \\
     \midrule
         Random Effects Estimates & $\mu$ \\
     \midrule
         Expert intercept $(_i)$ & .127 \\
         Hop intercept $(_j)$    & .000 \\
    \midrule
      Residual $\epsilon_{i,j}$ & .609 \\
    \bottomrule
    \multicolumn{5}{r}{N = 532, DF = 522, AIC = 1066.3, BIC=1121.7}
    \end{tabular}
\end{table}

\subsection{Evade Hops}
Table \ref{tab:evade_m} shows all effects retained in the final model with the outcome variable of overall evade hop difficulty (midpoints). Four fixed effects were retained. Experts rated a hop as less difficult to evade when more information is available to aid with this, but more difficult to evade when they were uncertain about the availability of such information. Overall evasion difficulty was also higher for hops relating to more mature technologies. A negative interaction term was evident for ratings of hop complexity, with certainty around a more complex hop being associated with it being more difficult to evade, but certainty around a hop being less complex associated with it being easier to evade. The availability of information was found to be have the most robust effect.

\begin{table}[t]
    \centering
    \caption{Results showing significant effects of hop attribute midpoints (m), widths (w) and two-way interactions ($m \cdot w$) on midpoints of overall evade hop difficulty ratings.\label{tab:evade_m}}
    \begin{tabular}{lrrrr}
    \toprule
    Fixed Effects Estimates & $\beta$ & SE & $t$ & $p$ \\
     \midrule
        Intercept : $(_0)$                                                         & -.023 & .133 & -.173  & .863   \\
        Availability Information $m$ : $(x^{am}_{i,j})$                                      & -.240 & .049 & -4.895 & $<.001$ \\
        Maturity $m$ : $(x^{rm}_{i,j})$                                          & .177  & .051 & 3.459  & $<.001$ \\
        Availability Information $w$ : $(x^{aw}_{i,j})$                                      & .142  & .049 & 2.878  & .004    \\
        Complexity $m \cdot w$ : $(x^{cm}_{i,j} \cdot x^{cw}_{i,j})$         & -.105 & .053 & -1.993 & .047    \\
    \midrule
         Random Effects Estimates & $\mu$ \\
    \midrule
         Expert intercept $(_i)$ & .457 \\
         Hop intercept $(_j)$    & .340 \\
     \midrule
      Residual $\epsilon_{i,j}$ & .772 \\
    \bottomrule
    \multicolumn{5}{r}{N = 418, DF = 413, AIC = 1081.8, BIC=1114.0}
    \end{tabular}
\end{table}

\begin{table}[t]
    \centering
    \caption{Results showing significant effects of hop attribute midpoints (m), widths (w) and two-way interactions ($m \cdot w$) on widths of overall evade hop difficulty ratings.\label{tab:evade_w}}
    \begin{tabular}{lrrrr}
    \toprule
    Fixed Effects Estimates & $\beta$ & SE & $t$ & $p$ \\
    \midrule
        Intercept : $(_0)$                                                & -.000 & .058 & -.000 & $>.999$ \\
        Complexity $w$ : $(x^{cw}_{i,j})$                                & .241  & .046 & 5.200 & $<.001$ \\
        Availability Information $w$ : $(x^{aw}_{i,j})$                             & .440  & .045 & 9.683 & $<.001$ \\
        Maturity $w$ : $(x^{rw}_{i,j})$                                 & .134  & .045 & 2.982 & .003    \\
        \midrule
         Random Effects Estimates & $\mu$ \\
     \midrule
        Expert intercept $(_i)$ & .070 \\
        Hop intercept $(_j)$    & .159 \\
    \midrule
      Residual $\epsilon_{i,j}$ & .643 \\
    \bottomrule
    \multicolumn{5}{r}{N = 418, DF = 414, AIC = 863.0, BIC=891.2}
    \end{tabular}
\end{table}

Table \ref{tab:evade_w} shows all effects retained in the final model with the outcome variable of uncertainty surrounding overall evade hop difficulty (widths). These results show that experts were more uncertain in their overall hop rating when they were more uncertain about each of the three attributes of a given hop: complexity, information availability, or maturity. However, no significant main effects of attribute rating position, nor any interaction terms were found. Uncertainty surrounding information availability was found to have the most robust effect.

\section{Conclusions\label{sec:conclusions}}
We analyse ratings provided by cyber-security experts that pertain to a range of potentially important component (hop) attributes, previously identified as commonly occurring within attack vectors of mainstream government cyber-systems. Importantly, these ratings were obtained through interval-valued responses, which enable experts to indicate both their rating and the uncertainty associated with this rating in a single, integrated response.

Our analyses provide a `proof of concept' for interval-valued data capture applied to the field of cyber-security. We identify key factors that contribute to both component vulnerability and uncertainty, depending on whether a hop requires compromising or only bypassing. For example, the availability of information has the largest impact on the overall difficulty of evading a component, while uncertainty around the inherent difficulty of an attack has the largest impact on its overall uncertainty.

Uncertainty in experts' attribute ratings is found to be valuable in determining not only overall uncertainty, but also overall hop vulnerability. In a number of specific cases, this information explained variance over and above the discrete midpoints of attribute ratings. For instance, when predicting the overall difficulty of attack hops, uncertainty around the maturity of technology of a given hop was associated with a significant increase in difficulty rating for that hop. In other cases, we found that in order to best explain overall difficulty ratings it was necessary to consider an interaction effect, between the position and width of responses. Sometimes, the combination of both factors provided a better predictor than either did alone. It was the novel use of an interval-valued response format that made it possible to coherently capture this uncertainty, alongside traditional ratings.

This study provides initial empirical evidence for the potential added-value offered by capturing interval-valued responses to model expert uncertainty. This is demonstrated in the case of modelling vulnerabilities within cyber-systems comprising multiple components, each with varying attributes, as is characteristic of cyber-physical systems. The benefit of using interval-valued responses was found using a comparatively low-complexity linear modelling approach. While such an approach is unsuited to capturing varying effects of responses along the response scale (for instance, it cannot account for a tendency for responses to saturate towards the extremities), its simplicity facilitates interpretation of the results. In future work, we will investigate the use of generalised additive models, within which the relationships between independent and dependent variables may be non-linear. Additionally, we are pursuing an ongoing programme of research to demonstrate the efficacy of interval-valued responses, both in terms of capturing uncertainty and improving predictive power, with reference to real-world ground-truth.

\newpage

\bibliographystyle{splncs04}

\end{document}